\documentclass[a4paper,11pt]{article}
\pdfoutput=1 

\usepackage{jinstpub} 

\usepackage{graphicx}
\usepackage{caption}
\usepackage{amsmath}
\usepackage{hyperref}
\usepackage{txfonts}
\usepackage{upgreek}
\usepackage{multirow}
\usepackage{subfig}

\usepackage{appendix}

\newcommand{\cpad}{\ensuremath{\mathrm{C_{pad}}}}
\newcommand{\cpi}{\ensuremath{\mathrm{C_{\uppi}}}}
\newcommand{\ph}{\ensuremath{\upgamma}}
\newcommand{\muon}{\ensuremath{\upmu}}

\title{A detector-emulation method for realistic readout-electronics tests.\\ A case study of VMM3a ASIC for sTGC detector.}

\author[a,1]{L. Moleri,\note{Corresponding author. Current affiliation: Weizmann Institute of Science}}
\author[a]{N. Lupu,}
\author[a]{A. Vdovin,}
\author[a]{and E. Kajomovitz,}

\affiliation[a]{Physics Department,
Technion - Israel Institute of Technology,\\
3200003, Haifa, Israel\\}

\emailAdd{m.luca@campus.technion.ac.il}

\abstract{We present a detector-emulation method for developing, optimizing, and testing front-end electronics for particle detectors under realistic experimental conditions. The method is capable of reproducing bipolar signals shape and amplitude, rates, pileup, and radiation backgrounds. By controlling the different components of the experimental environment, the method allows for the assessment of their individual and combined effects, which is difficult to achieve in test beam or experiment environment.

The technique emulates the output current of AC coupled detector terminals for an experimental environment of interest, by combining a sequence of recorded or simulated signal waveforms. Voltage waveforms produced with off the shelf arbitrary waveform generators are converted to current by a simple circuit, which also accounts for the characteristic capacitance of the detector, and injected into the front-end electronics.
We demonstrate the technique by applying it in the optimization and characterization of the  front-end electronics of the sTGC detector for the ATLAS experiment. 
}


\begin{document}
\maketitle
\flushbottom

\section{Introduction}
\label{sec: intro}

Coupling a detector to a readout ASIC requires an evaluation of their performance as a system in a specific experimental environment. 
First of all the choice of the ASIC and the optimization of its adjustable parameters (e.g. gain, sensitivity and shaping or integration time) are determined by the characteristic detector output-current time development (signal shape).
In the particular case under study in this work an optimization to match the signal from the sTGC pads to the FE ASICS dynamic range is necessary i.e. to minimize dead-time given the high amplification of the detector, and the large charge that can be deposited by the background radiation. 
For these purposes, the traditional pulse injection method - a square wave through a capacitor- is not resembling a realistic detector behavior. 
Moreover, the detector-ASIC system should be evaluated in the relevant radiation environment for an application, including signal and background sources. While the ultimate test is typically done on a fully instrumented prototype in an accelerator particle beam, preliminary stages of the R$\&$D are usually performed in a laboratory using various radiation sources, including cosmic muons. This can be complicated and expensive and often insufficient.

The method presented in this work relies on a sequence of recorded or simulated signal waveforms, which makes it independent on the availability of a real detector and radiation source. 
Voltage waveforms produced with of-the-shelf arbitrary waveform generators are converted to current by a simple circuit. We employed the common scheme of "charge injection through a capacitor" to develop a more general detector emulation method that can mimic the realistic output current of AC coupled detector terminals. 
The bipolar nature of the produced current is intrinsic for example to gaseous detectors which include resistive layers (TGC cathodes~\cite{smakhtin2009thin}, resistive Micromegas~\cite{alexopoulos2011spark}, RPC~\cite{SANTONICO1981377} and others). On the other hand, in some detector concepts the readout terminals are also charge collectors, so that the signal induced on them is purely unipolar (e.g. GEM~\cite{guedes2003effects}, THGEM~\cite{bhattacharya2019signal}, SiPM~\cite{corsi2007modelling}). In this case the method would still work, but one should be aware that effects on the ASIC performance due to non-zero DC currents could be missed (e.g. baseline shift, saturation).

It should be noted that detector terminals are always AC coupled to the readout electronics through a decoupling capacitor when they have to be biased at a voltage different than the ASIC input working point. This is the case for example for TGC anode wires~\cite{smakhtin2009thin}.

A vast set of detector and signal source features can be reproduced by the emulator, in particular signal shape and amplitude, detector capacitance, source rates, pileup, and radiation backgrounds. Their individual and combined effects on the system performance can be assessed, which can hardly be done using a real detector experiment, even in accelerator beam tests.

The method is presented here and applied to a specific case in the context of the New Small Wheel (NSW)~\cite{stelzer2016new}, which is part of the upgrade of the ATLAS muon-spectrometer~\cite{palestini2003muon}. The NSW system includes two different detector technologies: the small-strip Thin Gap Chamber (sTGC)~\cite{smakhtin2009thin}, which is a multi-wire proportional chamber, and the Micromegas~\cite{wotschack2013development}, which is a micro-pattern gaseous detector. The two technologies have very different gains at the nominal operation voltages, specifically, the sTGC signal can be $\sim$10 times larger than the Micromegas one. A common ASIC was developed for the readout of both detectors: the VMM3a chip~\cite{iakovidis2018vmm}. We studied the sTGC detector coupling to the readout chip.
In particular we focused on the optimization of the signal-to noise and the readout dynamic range, namely, maximizing the detection efficiency and minimizing the number of events saturating the chip analog output which also results in undesirable dead time.

Since the detector output (rate, amplitude, etc) is set by the experimental environment and the nominal operation voltage in the experiment, we investigated different attenuator circuits and VMM3a configuration parameters. 

\section{The detector emulation method}
\label{sec: detector emulation}

The proposed detector emulation method aims at producing a realistic detector current signal as input for testing readout electronics. In particular, the signal features of the detector signal that should be emulated are its shape (positive/negative currents, time development) amplitude distributions, and expected rates.
For this purpose, we used a two-channel current-injection circuit (injector), depicted in figure~\ref{subfig: injector circuit}. The injector operation can be described by the formula $i(t) \sim \frac{d}{dt}V(t)$, where V(t) is the voltage at the injector input, the time derivative is the injector operation, and i(t) is the injector output current entering the readout electronics. The injector is in fact operating on the voltage pulse at its input approximately as a mathematical derivative. The additional attenuation circuit ($\uppi$-network) introduces an AC coupling, which does not significantly affect the current shape, as demonstrated by the LTspice simulation results in figure~\ref{subfig: LTspice simulation} for different injector capacitor values. 

The emulated detector-current signal at the injector output - i(t) - is obtained when the injector input fed with a sequence of voltage pulses V(t) which shape is proportional to the time integral of a real detector-current signal. The latter can be obtained by measurement, by calculation or by simulation, at the desired experimental conditions for the radiation source and detector.

In the present study, V(t) was provided by the measurement of cosmic muon signals from an sTGC detector, recorded with a charge sensitive pre-amplifier with long decay time, effectively operating as a current integrator. The details are given in appendix~\ref{app: det emulator input}.

\subsection{Experimental setup}
\label{subsec: experimental setup}

The injector input waveforms were recorded and played by two Arbitrary Waveform Generator (AWG) instruments, \href{https://www.caen.it/products/dt5800/}{CAEN DT5800} and \href{https://www.keysight.com/il/en/products/arbitrary-waveform-generators/m8100-series-arbitrary-waveform-generators.html}{Keysight AWG M8190A}, each of them connected to a separate injector channel. This allowed for injecting a mix of two kinds of emulated events, mimicking muons (\muon) and energetic photons (\ph) with controllable rates. The CAEN module is capable of producing waveforms with a custom shape, random pulse height drawn from a custom distribution, and random times (exponential time distribution) at a set average rate. The instrument has 16 memories available for signal pileup, providing a maximum rate of $\sim$1/(16$~\times~signal~duration$). For the used signal shape (see appendix~\ref{app: det emulator input}) the rate limit was $\sim$40~kHz. The CAEN DT5800 was used to produce both $\upmu$ and $\upgamma$ waveforms which differed in their amplitude distributions (spectra), whereas the signal shape was the same for both cases. The injected pulse height spectra were set to reproduce the ones measured with a sTGC detector at CERN-SPS test beam in GIF++ \ph-irradiation facility. $\upmu$-like signals were injected directly by the CAEN AWG at 1~kHz constant rate, in an injector channel. The $\upgamma$ signals were reproduced by the Keysight AWG at random times (exponential) at different average rates up to $\sim$1~MHz. To create the high $\upgamma$ rate sequences, the following method was used: several repetitions of $\upgamma$-like signals at random times were produced with the CAEN module at the maximum possible average rate, recorded by a scope and summed point by point in time to pile them up, effectively increasing the signal rate.

The VMM3a ASIC was installed onto a GP-VMM board~\cite{gkountoumis2018prototype}, from which it is possible to readout all chip outputs, analog and digital. Analog outputs and the 1-bit digital output ("direct output") were monitored and recorded from an oscilloscope, while the 10-bit ADC output was recorded from a PC by the VERSO DAQ software~\cite{verso}.
The parameters of the VMM3a chip were fixed to 3~mV/fC gain at 50~ns shaping time.

\begin{figure}[htbp]
\centering
\subfloat[Electronics schematics including the injector]{
\includegraphics[scale=0.4]{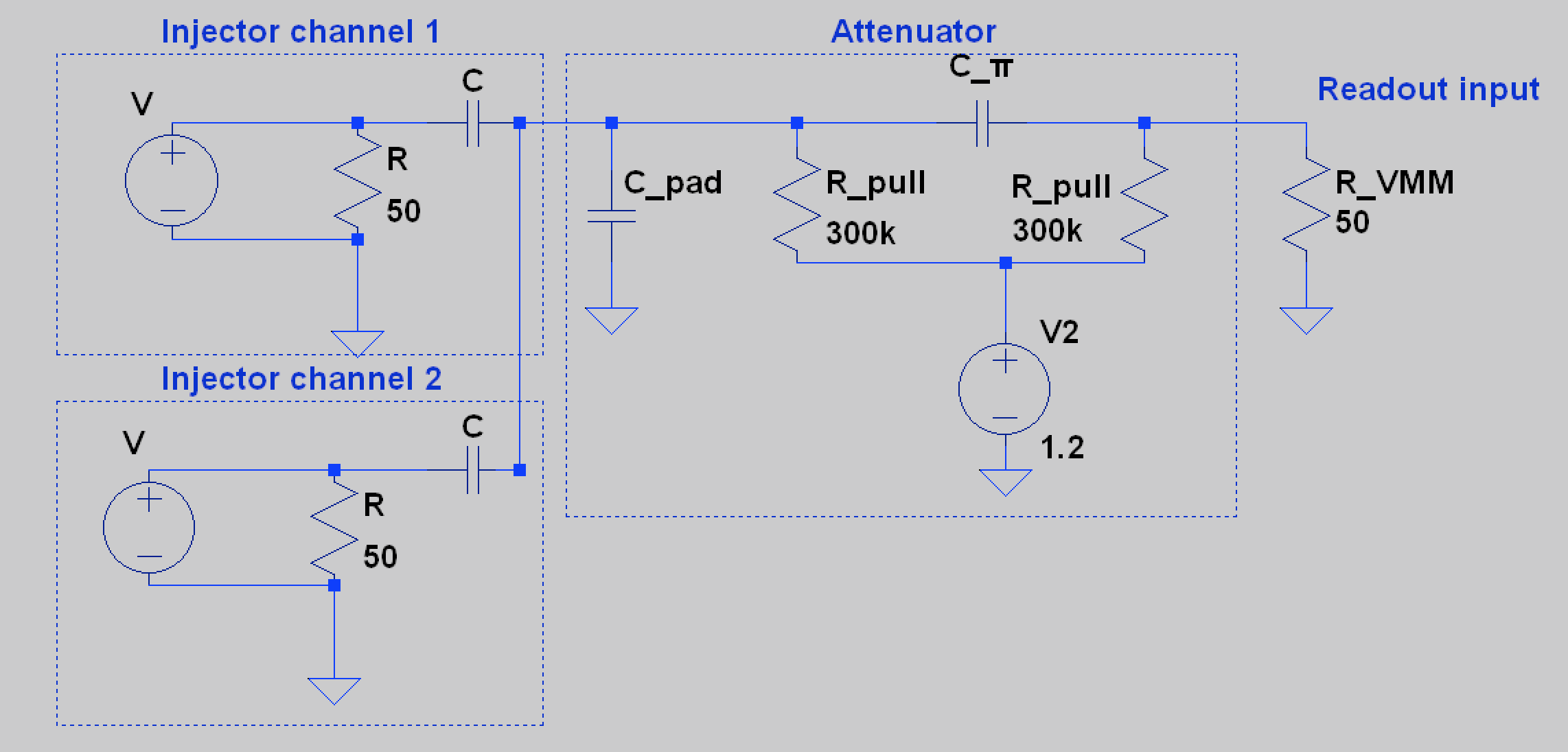}
\label{subfig: injector circuit}
}
\\
\subfloat[LTspice simulation of the injector operation]{
\includegraphics[scale=0.6]{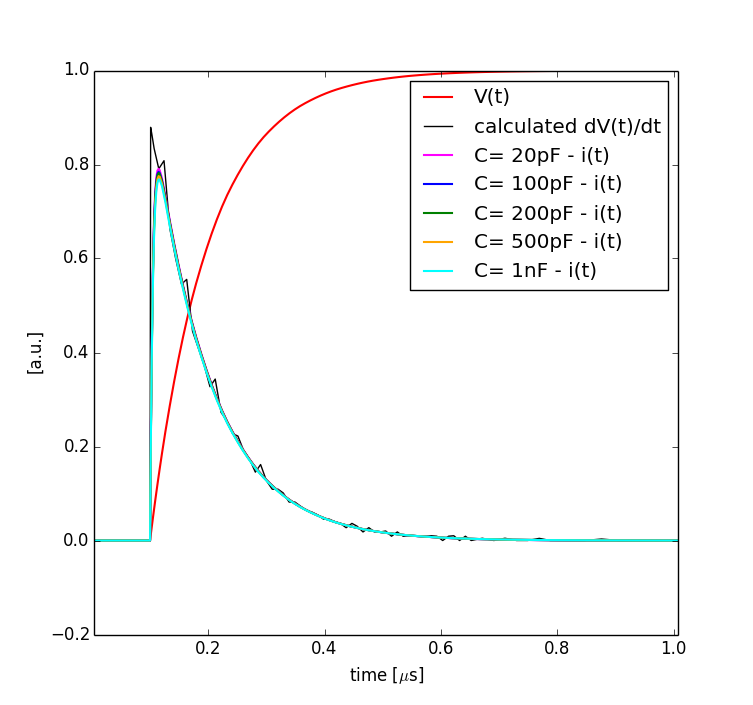}
\label{subfig: LTspice simulation}
}
\caption{\protect\subref{subfig: injector circuit} Electric scheme of the double current injector. Each of the two injector channels works independently: a voltage pulse V(t) is injected into a capacitor (C), which is connected through a $\uppi$-network attenuation circuit to the ASIC input R$\mathrm{_{VMM}}$. A capacitor (\cpad) emulates the capacitance of an sTGC pad. Both values of \cpad~and \cpi~can be changed. C= 100~pF.
\protect\subref{subfig: LTspice simulation} An LTspice simulation shows an exponential input V(t), together with normalized current pulses through R$\mathrm{_{VMM}}$ for a wide range of C values, with \cpad= 800~pF and \cpi= 100~pF. The calculated dV(t)/dt is also plotted, demonstrating that the expected injector output is well reproduced for a large range of C values.  
}
\label{fig: injector}
\end{figure}

The emulator method was applied to characterize the ASIC response in the presence of an external $\uppi$-network attenuation circuit, depicted in figure~\ref{subfig: injector circuit}. In order to mimic the real sTGC pad capacitance, a capacitor ($\mathrm{C_{pad}}$) was introduced in the $\uppi$-network scheme. The tested \cpad~ values were 800~pF and 2~nF (representative values for real sTGC modules). Table~\ref{tab: parameter space} summarizes the investigated capacitance values and the theoretical attenuation factor for each of their combinations, according to the formula $f_{att}= \cpi/(\cpi+\cpad)$. The injector capacitor value C used was 100~pF.

\begin{table}[h]
\begin{center}
\begin{tabular}{| c | c | c |}
\hline
\textbf{\cpad~[pF]} & \textbf{\cpi~[pF]} & \textbf{attenuation factor}\\ \hline
\multirow{4}{*}{800} & 100 & 0.1 \\ \cline{2-3}
& 200 & 0.2 \\ \cline{2-3}
& 300 & 0.27 \\ \cline{2-3}
& 470 & 0.37 \\ \hline
\multirow{5}{*}{2000} & 200 & 0.9 \\ \cline{2-3}
& 470 & 0.19 \\ \cline{2-3}
& 600 & 0.23 \\ \cline{2-3}
& 800 & 0.29 \\ \cline{2-3}
& 1000 & 0.33 \\ \hline
\end{tabular}
\caption{The explored detector parameter space.}
\label{tab: parameter space}
\end{center}
\end{table}

\section{Measurements and results}
\label{sec: detector emulator measurements and results}

The investigated parameter space consisted of the different detector and attenuator capacitance's values reported in table~\ref{tab: parameter space}, as well as different VMM3a configuration bits activated. Two kinds of measurements were performed: signal spectra for pure \muon-signal without \ph~background, and \muon-detection efficiency with \ph~background. 

\subsection{Parameter space reference point and emulated spectra}
\label{subsec: emulated spectra}

\begin{figure}[htbp]
\centering
\subfloat[\muon~ spectrum from test beam]{
\includegraphics[scale=0.55]{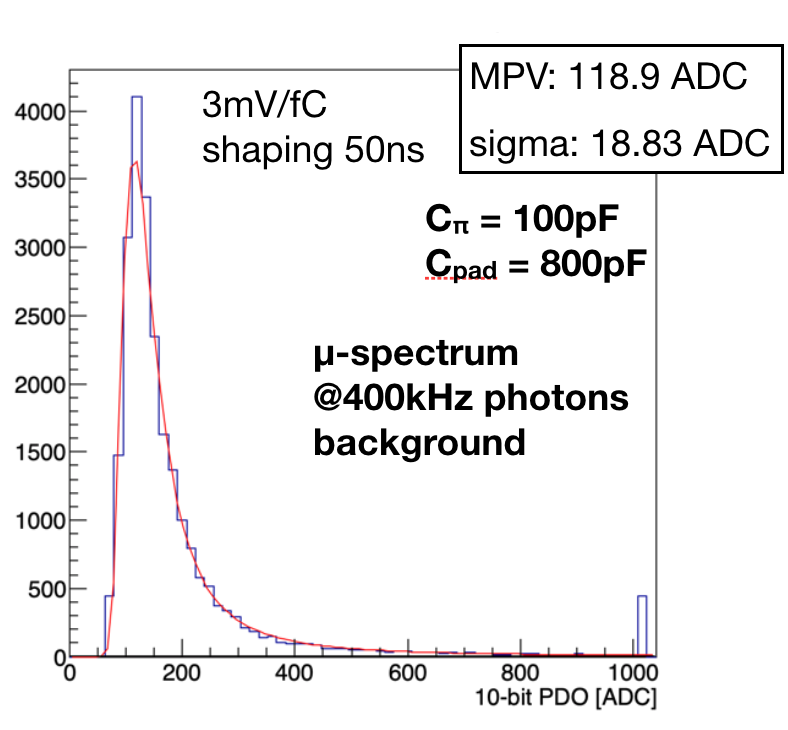}
\label{subfig: starting point TB mu}
}
\subfloat[Emulated \muon~ spectrum]{
\includegraphics[scale=0.55]{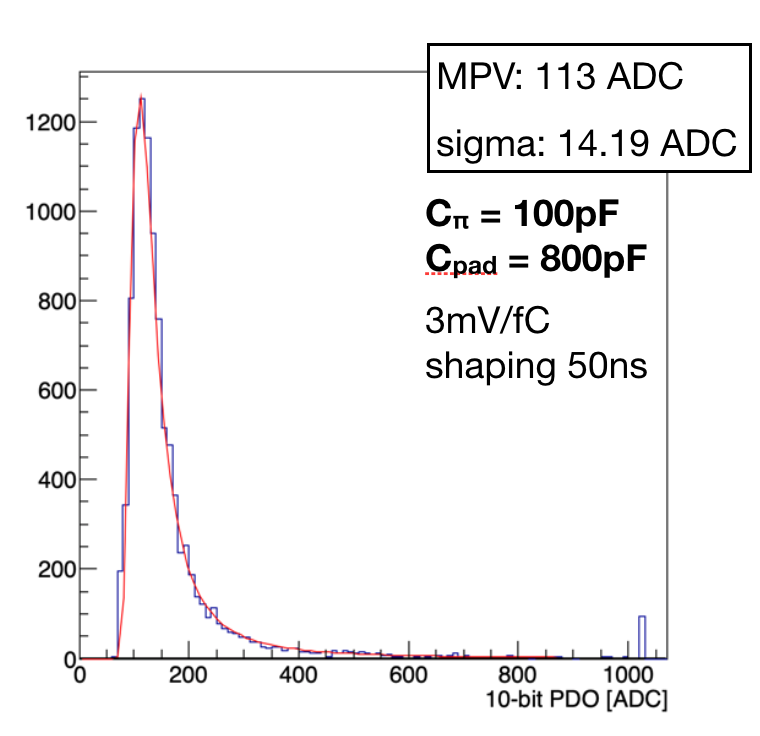}
\label{subfig: starting point emu mu}
}
\caption{VMM3a 10-bit ADC readout: \protect\subref{subfig: starting point TB mu} \muon~spectrum recorded from QL1 sTGC\cite{pudzha2020small} operated at 2900~V in GIF++ under $\sim$400~kHz \ph~background rate. \protect\subref{subfig: starting point emu mu} Emulated \muon~ spectrum used as reference point in the parameter space.}
\label{fig: starting point mu}
\end{figure}

The emulator circuit configuration with \cpad= 800~pF, and \cpi= 100~pF, was taken as the reference point in the parameter space to tune the AWG spectra.
We set the emulator \ph~ and \muon~ pulse height distributions to reproduce those measured in GIF++ at with a sTGC chamber (type QL1) operated at 2900~V under \ph~ rate of $\sim$400~kHz in the considered pad. In this way, the voltage drop (i.e. gain drop) caused by the current flowing through the sTGC wires is taken into account for a reasonably high \ph~rate. The emulated \muon~spectrum reproduces well the original one, as shown in figure~\ref{fig: starting point mu}. In order to obtain the correct fraction of saturated events, the tail of the emulated charge distribution was cut at 12~times the Landau distribution most probable value (MPV).

\begin{figure}[htbp]
\centering
\subfloat[\muon~ spectra]{
\includegraphics[scale=0.35]{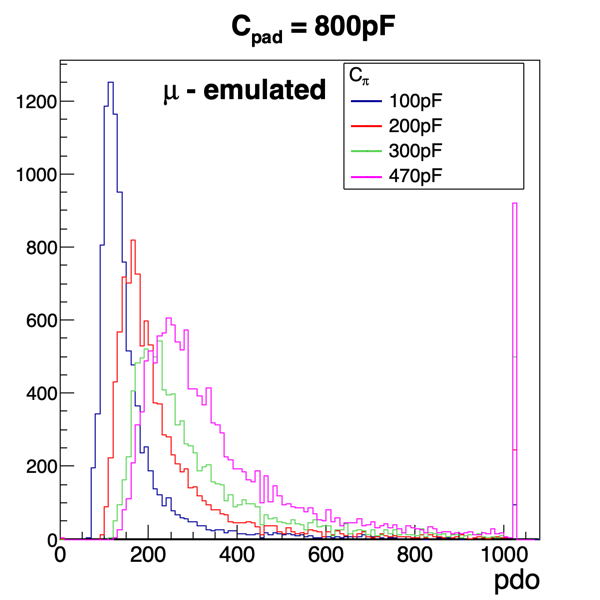}
\label{subfig: spectra mu}
}
\subfloat[\ph~ spectra]{
\includegraphics[scale=0.35]{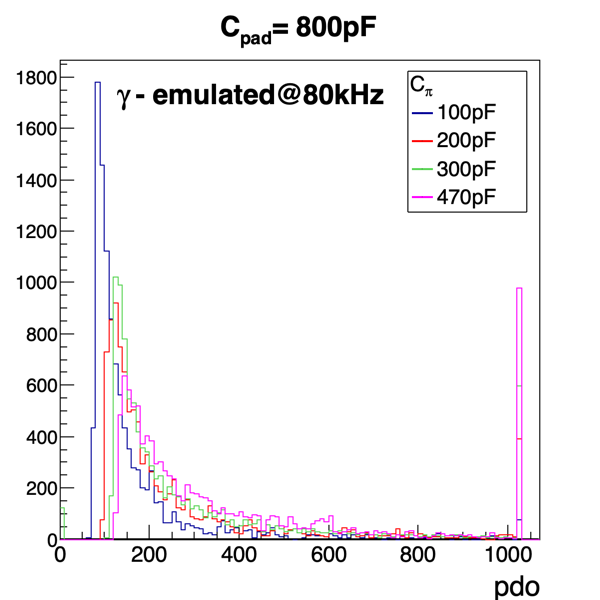}
\label{subfig: spectra ph}
}
\caption{VMM3a 10-bit ADC readout: Emulated \muon~(\protect\subref{subfig: spectra mu}) and \ph~(\protect\subref{subfig: spectra ph}) spectra for different values of \cpi, for \cpad= 800~pF.
}
\label{fig: spectra emulated}
\end{figure}

After setting the reference point, the parameter space was explored by changing the values of \cpad~and \cpi~and the VMM3a bits.
The \muon~and \ph~spectra obtained with different \cpi~values are presented in figure~\ref{fig: spectra emulated}, for \cpad= 800~pF. It can be seen that the most probable value (MPV) of the $\upmu$~ distributions changes according to the attenuation factors summarized in table~\ref{tab: parameter space}.

\subsection{Efficiency measurements}
\label{subsec: efficiency measurement}

\begin{figure}[htbp]
\centering
\includegraphics[scale=0.5]{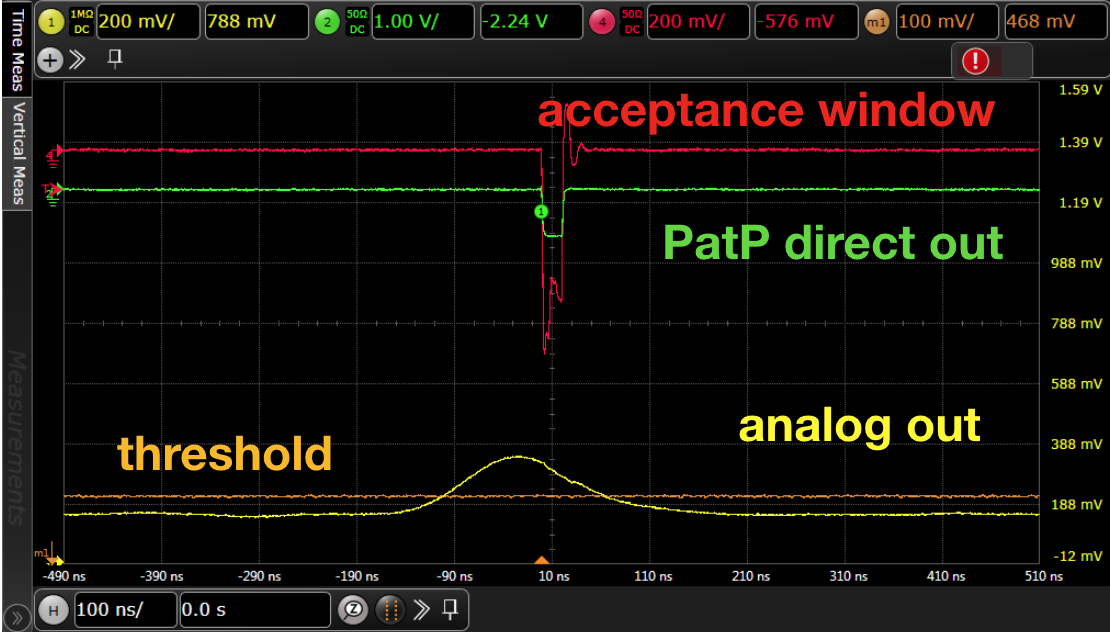}
\caption{Example of efficient \muon~event: the analog VMM3a signal (Yellow) passes the threshold (Orange), causing a pulse at the peak time in the 1-bit direct output (Green). The coincidence between the 1-bit output and the acceptance window (Red) determines the efficiency of the event.
}
\label{fig: efficiency definition}
\end{figure}

The effect of the \ph~background rate on the \muon~detection efficiency was tested at different points of the parameter space.
The \muon~detection efficiency was defined by means of a 20~ns acceptance window, generated simultaneously with every \muon~event.
The coincidence between the VMM3a 1-bit direct output in pulse-at-peak (PatP) mode and the acceptance window determines the efficiency of the event. The relative delay was adjusted so that the VMM3a 1-bit pulse fell, on average, in the middle of the time window. Since the PatP mode provides a $\sim$ 25~ns long pulse, its overlap with the 20~ns time windows gives an effective 45~ns time window. 
The VMM3a digital output is ready to accept new signals only after an idle time caused by re-arming (i.e. analog signal returning below threshold) plus a fixed digitization time which is measured to be $\sim$60~ns and is increased to $\sim$200~ns in case that the 10-bit digitization is also active.  
An example of efficient isolated \muon~ event is depicted in figure~\ref{fig: efficiency definition}. 

\begin{figure}[htbp]
\centering
\subfloat[Pileup]{
\includegraphics[scale=0.5]{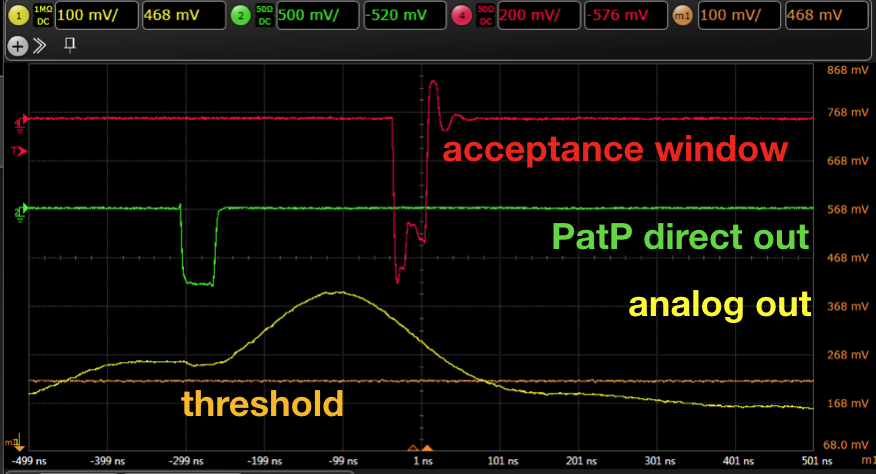}
\label{subfig: cause1 pileup}
}
\subfloat[Jitter]{
\includegraphics[scale=0.58]{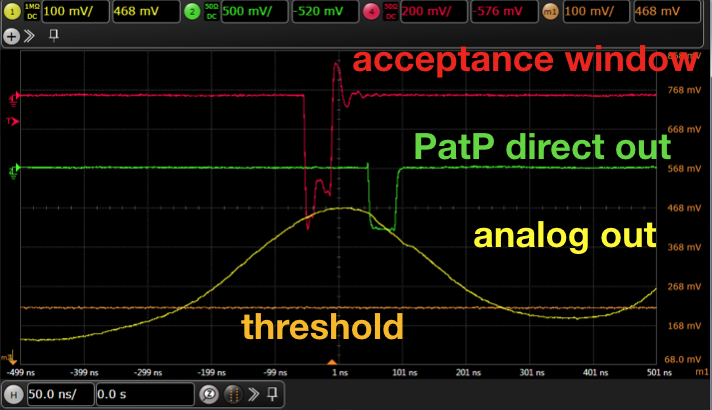}
\label{subfig: cause2 jitter}
}

\subfloat[Deep saturation]{
\includegraphics[scale=0.6]{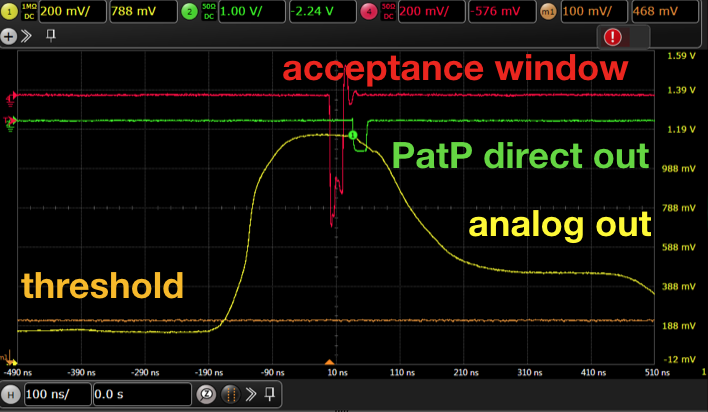}
\label{subfig: cause3 deepsat}
}
\caption{Possible causes of \muon~ detection inefficiency.
}
\label{fig: inefficiency causes}
\end{figure}

In the presence of the \ph-background, the picture gets more complicated, and more sources of inefficiency (other than signal below threshold) come into play. A collection of possible causes of inefficiency is given in figure~\ref{fig: inefficiency causes}. When a \ph~comes just before a \muon~ and the two induced signals merge (pileup) so that the voltage does not go below threshold, the event is inefficient because the VMM3a digitizer is not re-armed (figure~\ref{subfig: cause1 pileup}). An event can also be inefficient because of jitter. In the present emulation the intrinsic sTGC signal jitter is not included whereas there is a jitter component caused by the \ph~background (figure~\ref{subfig: cause2 jitter}). Other than jitter, the digital output and the acceptance window can be mismatched because of deep saturation of the analog signal, causing the peak to be detected at the wrong time (figure~\ref{subfig: cause3 deepsat}). It is also important to note that deeply saturated signals have a long tail, which can remain above the threshold for several $\upmu$s (figure~\ref{subfig: cause3 deepsat}). The time extension of the tail increases with the amount of charge in the pulse.

From this description it is clear that the number of deeply saturated events is a cause of severe inefficiency. Within the linear range of the chip, instead, the spectrum shape affects the efficiency mainly because of the time over threshold (ToT) and consequent rearming time. In particular, in case the spectrum is scaled by a factor, without changing the threshold, the ToT will increase. For this reason, in order to minimize the ToT effect and to have a fair comparison between measurements with different attenuation values, the threshold should always be set at a fixed value with respect to the spectrum shape (e.g. at the beginning of the distribution rising edge). Here, the threshold for each point in the parameter space was set so that the measured \muon~ efficiency with no \ph~ background was 98$\%$. 

\begin{figure}[htbp]
\centering
\subfloat[]{
\includegraphics[scale=0.35]{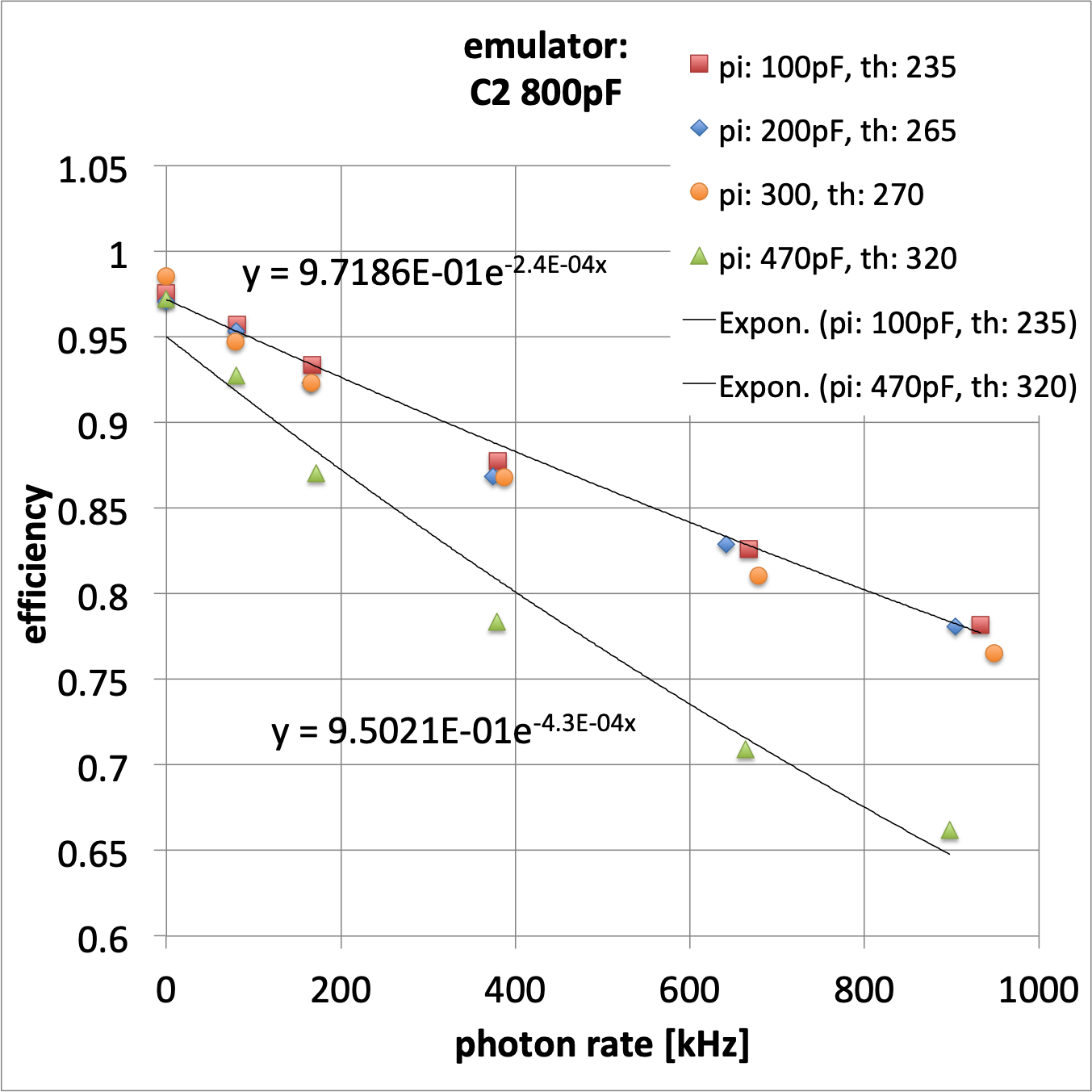}
\label{subfig: eff_QL1}
}
\subfloat[]{
\includegraphics[scale=0.35]{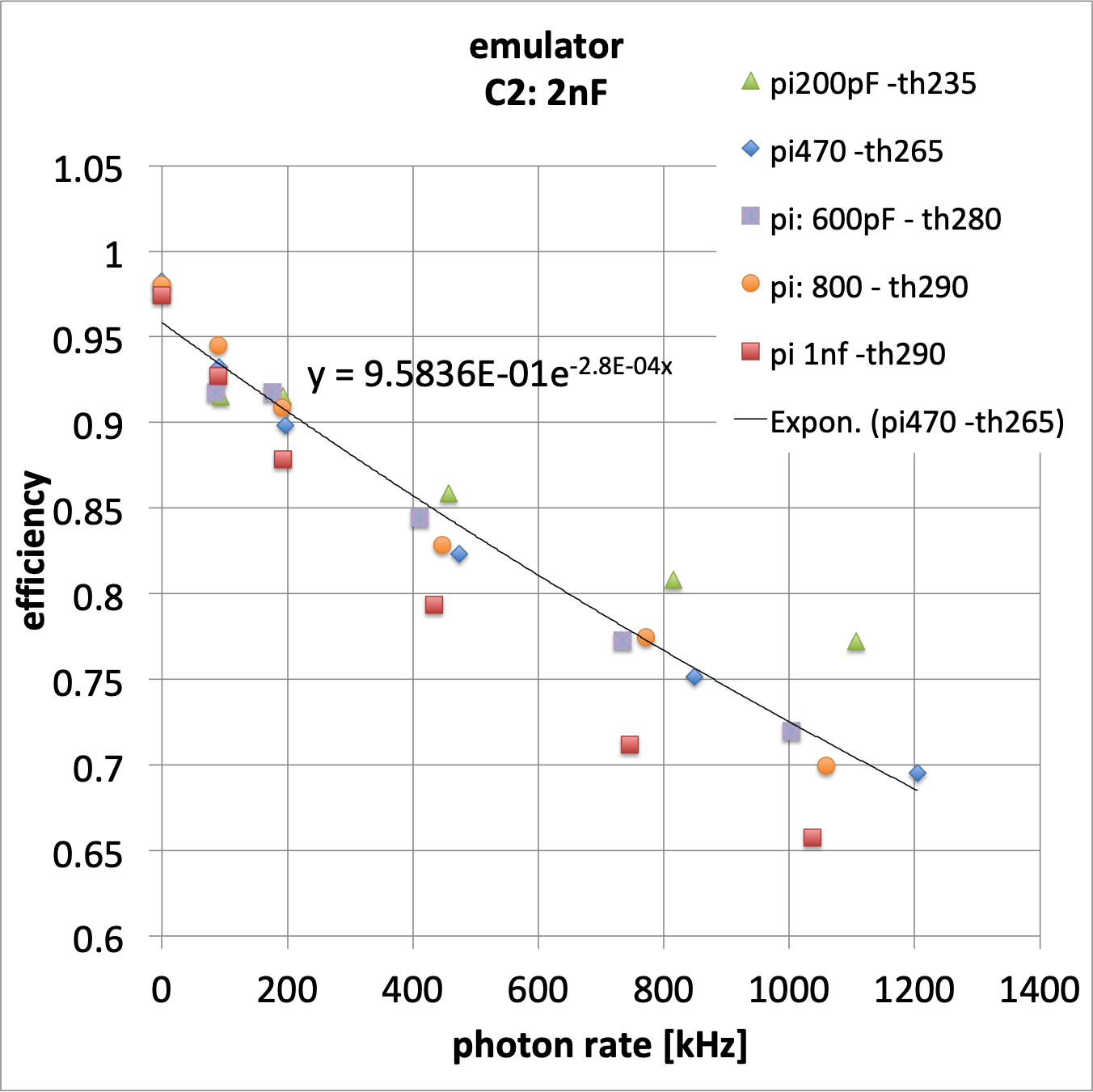}
\label{subfig: eff_QS3}
}
\caption{Emulated \muon~detection efficiency as a function of the \ph~background rate for different values of \cpi~, for \cpad= 800pF \protect\subref{subfig: eff_QL1} and 2~nF \protect\subref{subfig: eff_QS3} values. Exponential fits are also shown.
}
\label{fig: efficiency_results}
\end{figure}

\begin{figure}[htbp]
\centering
\includegraphics[scale=0.35]{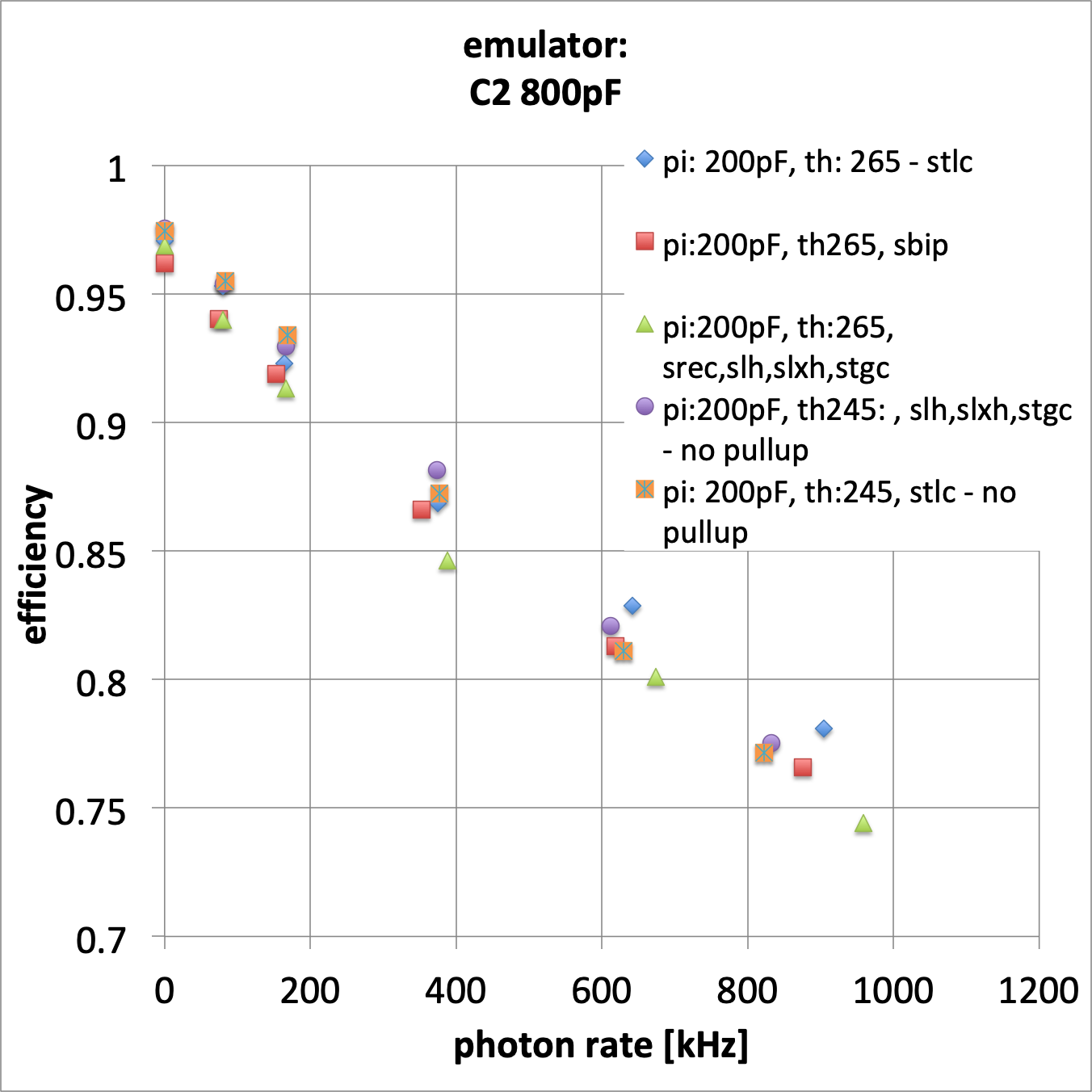}
\caption{Emulated \muon~ detection efficiency as a function of the \ph~background rate for different VMM3a configuration bits and pull-up currents, for \cpad= 800~pF prototype values.
}
\label{fig: efficiency bits}
\end{figure}

The results of the efficiency measurements are presented in figure~\ref{fig: efficiency_results}. The photon rate was obtained by counting the rate of VMM3a 1-bit outputs divided by the measured inefficiency. It can be seen that for both \cpad~values of 800~pF and 2~nF, the efficiency curve does not change significantly for attenuation factors larger than $\sim$0.3 (table~\ref{tab: parameter space}). This strengthens the hypothesis that the fraction of deeply saturated events is the main cause of inefficiency whereas in the linear range the details of the spectrum shape are less important, as long as the spectrum is well above noise. Since the time between two hits is random, it is possible to extract from the data the average intrinsic dead time by an exponential fit of the efficiency curves (see appendix~\ref{app: efficiency formula}). The result is $\sim$240~ns and $\sim$280~ns for \cpad~of 800~pF and the 2~nF respectively. 
Measurements with different VMM3a bits configurations are shown in figure~\ref{fig: efficiency bits}. In the same plot we show a measurement without pull-up current (V2 in the scheme in figure~\ref{subfig: injector circuit} was disconnected). In all the cases tested there was no appreciable difference in the measured efficiency trend.

\section{Monte Carlo simulation}
\label{sec: detector emulator MC}

\begin{figure}[htbp]
\centering
\includegraphics[scale=0.35]{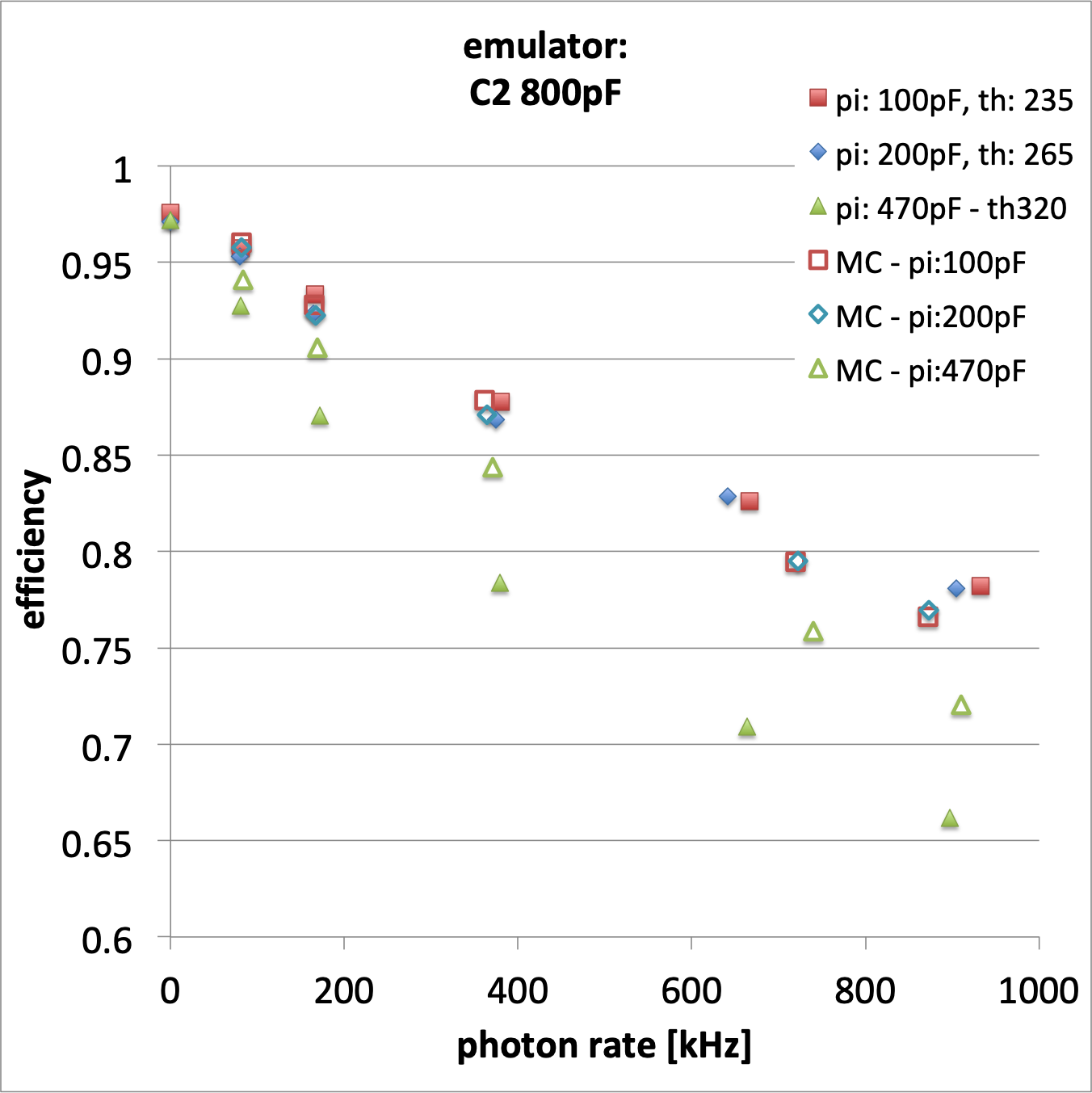}
\caption{The comparison between emulated and MC results for the \cpad= 800~pF configuration with different \cpi~ values.
}
\label{fig: efficiency MC}
\end{figure}

To validate the effect of the different sources of inefficiency (see section~\ref{subsec: efficiency measurement}), we compared efficiency measurements with a Monte Carlo (MC) simulation. The inputs to the simulation were the measured ToT distribution and the \ph~and \muon~spectra. The signal jitter was considered negligible. The simulation was carried out by generating photon events at random times. Each event caused a dead time t$\mathrm{_{dead}}$= ToT + 60~ns. If another signal was generated during dead time, the latter would be extended accordingly. The \muon~efficiency was calculated by checking the state of the system at a rate of 1~kHz. If the check occurred during dead time, the event would be considered inefficient.
The comparison between emulated and MC results are presented in figure~\ref{fig: efficiency MC} for the \cpad= 800~pF configuration with different \cpi~values.
The MC simulation reproduces the results quite well, indicating that we could identify the main causes of inefficiency in the linear range of the ASIC (see section~\ref{subsec: efficiency measurement}). The emulated inefficiency for \cpi~= 470~pF could not be well reproduced, supporting the hypothesis that it is caused mainly by deeply saturated events, which were not included in the MC simulation. 

\section{Conclusion}
\label{sec: conclusion}

The proposed detector emulator method is demonstrated to be a powerful tool to produce realistic output current signals for detectors having AC coupled terminals. 
This allows to characterize different combinations of detectors and readout electronics for an experimental environment of interest.
We reproduced the case of the VMM3a ASIC reading out sTGC cathode like signals. The emulated \muon~ and \ph~ spectra reproduced correctly the measured ones, allowing to predict the effect of different signal attenuations, to best exploit the ASIC dynamic range. We could also study the \muon~detection efficiency under increasing \ph~background rate. Different sources of inefficiency were recognized. In particular, the chip dead time caused by time-over-threshold could be equalized for different attenuation values by a proper threshold setting. The fraction of saturated signals seems to constitute a major cause of dead time. This interpretation is supported by a Monte Carlo simulation. 

We showed that the present method is a valuable testing tool, but still it has some limitations and caveats. In case that the emulator signal input is obtained from a charge sensitive preamplifier- like here-, it can be that some artificial currents are produced for baseline restoration (see appendix~\ref{app: det emulator input}). A possibility to avoid this would be using a simulated input signal.

More fundamentally, there are limits due to the AC coupling of the injector scheme to the detector. The pileup of the input voltage pulses can saturate the signal generator output, limiting the maximum rate of events. Moreover, true unipolar currents cannot be injected at all. Therefore we suggest a possible development of the method.
By injecting current through a resistor (instead of a capacitor), and using an input voltage waveform proportional to the current expected from the detector (instead of its integral), the emulator method can be extended to detectors with DC coupled readout terminals. The injection through a resistor would enable the emulation of bipolar and true unipolar current signals, and has the aditional advantage of allowing the emulation of a greater signal pileup than the capacitor based injector. It must be noted that this scheme would not work if the injector DC bias is different from the ASIC input working point. In this case the ASIC working point could be maintained by introducing a large blocking capacitor in series with the resistor -which would again exclude the possibility of injecting true unipolar currents.
\appendix
\section{Detector emulator signal input: acquisition and processing}
\label{app: det emulator input}

\begin{figure}[htbp]
\centering
\subfloat[All signals]{
\includegraphics[scale=0.22]{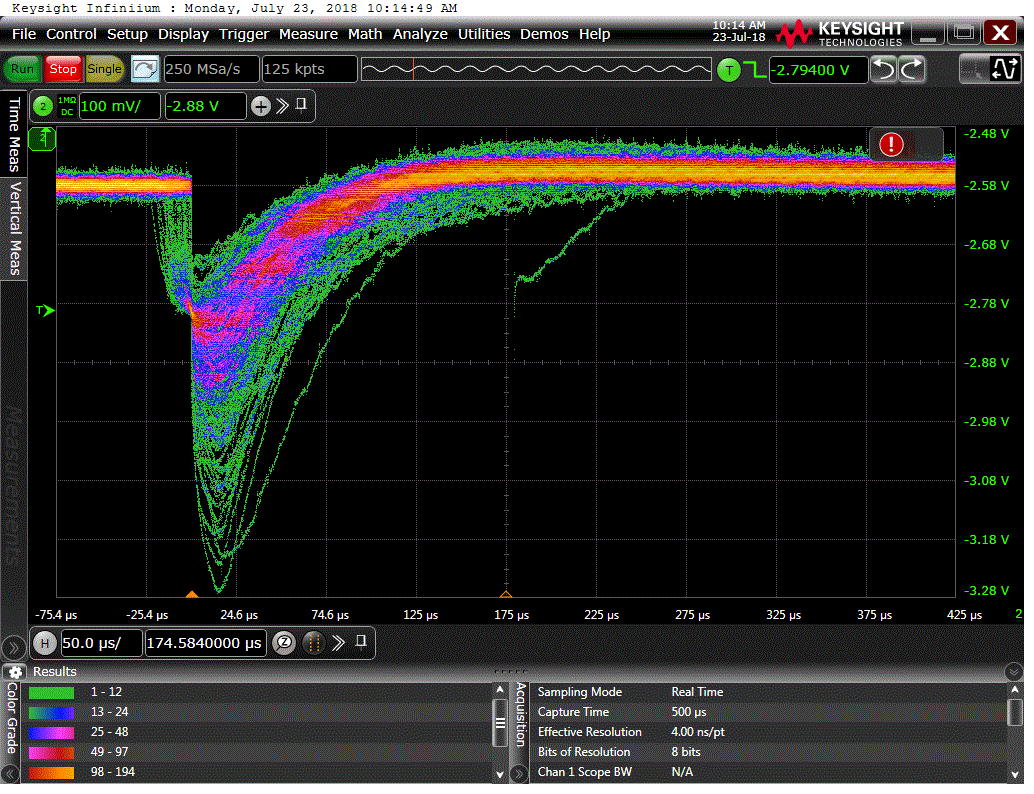}
\label{subfig: all signals}
}
\subfloat[Fast signals selection]{
\includegraphics[scale=0.22]{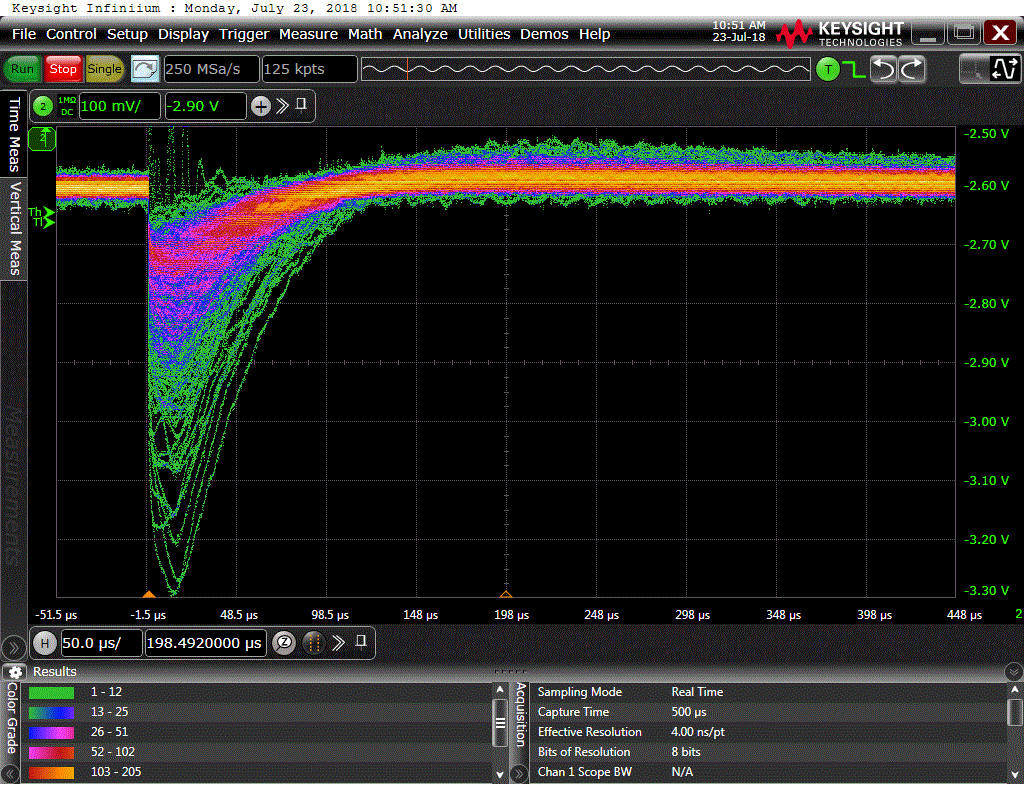}
\label{subfig: fast signals}
}
\caption{\protect\subref{subfig: all signals} Population of cosmic $\upmu$ signals recorded by sTGC detector pad in the standard CO$_2$/(45\%)n-pentane mixture at 2900~V, from a charge sensitive preamplifier. Notice the two populations of direct and crosstalk signals. \protect\subref{subfig: fast signals} The signals population after trigger selection of direct signals.
}
\label{fig: sTGC signals population}
\end{figure}

\begin{figure}[htbp]
\centering
\subfloat[Complete signal]{
\includegraphics[scale=0.5]{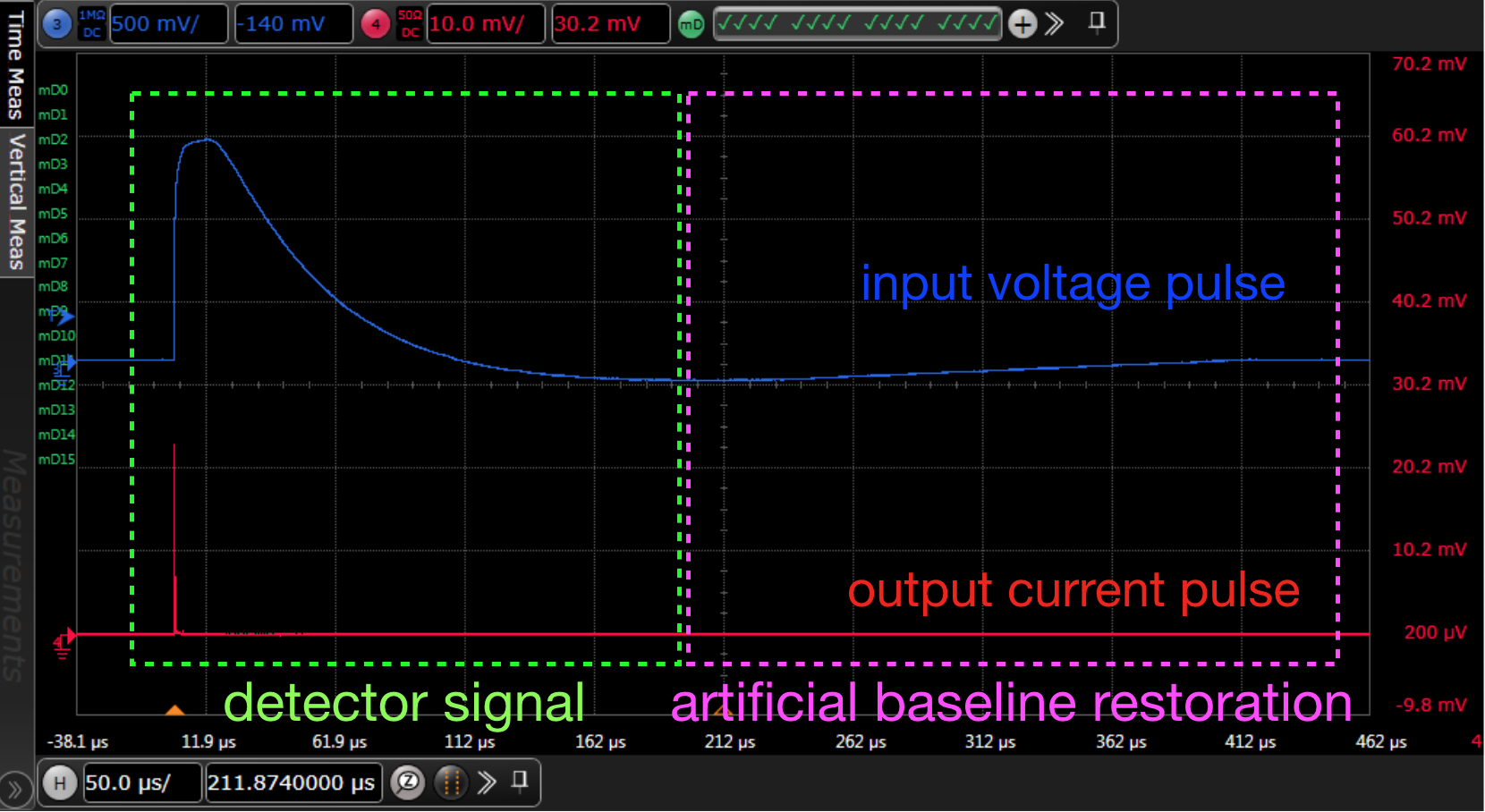}
\label{subfig: complete signal}
}
\\
\subfloat[Zoom in positive detector current]{
\includegraphics[scale=0.52]{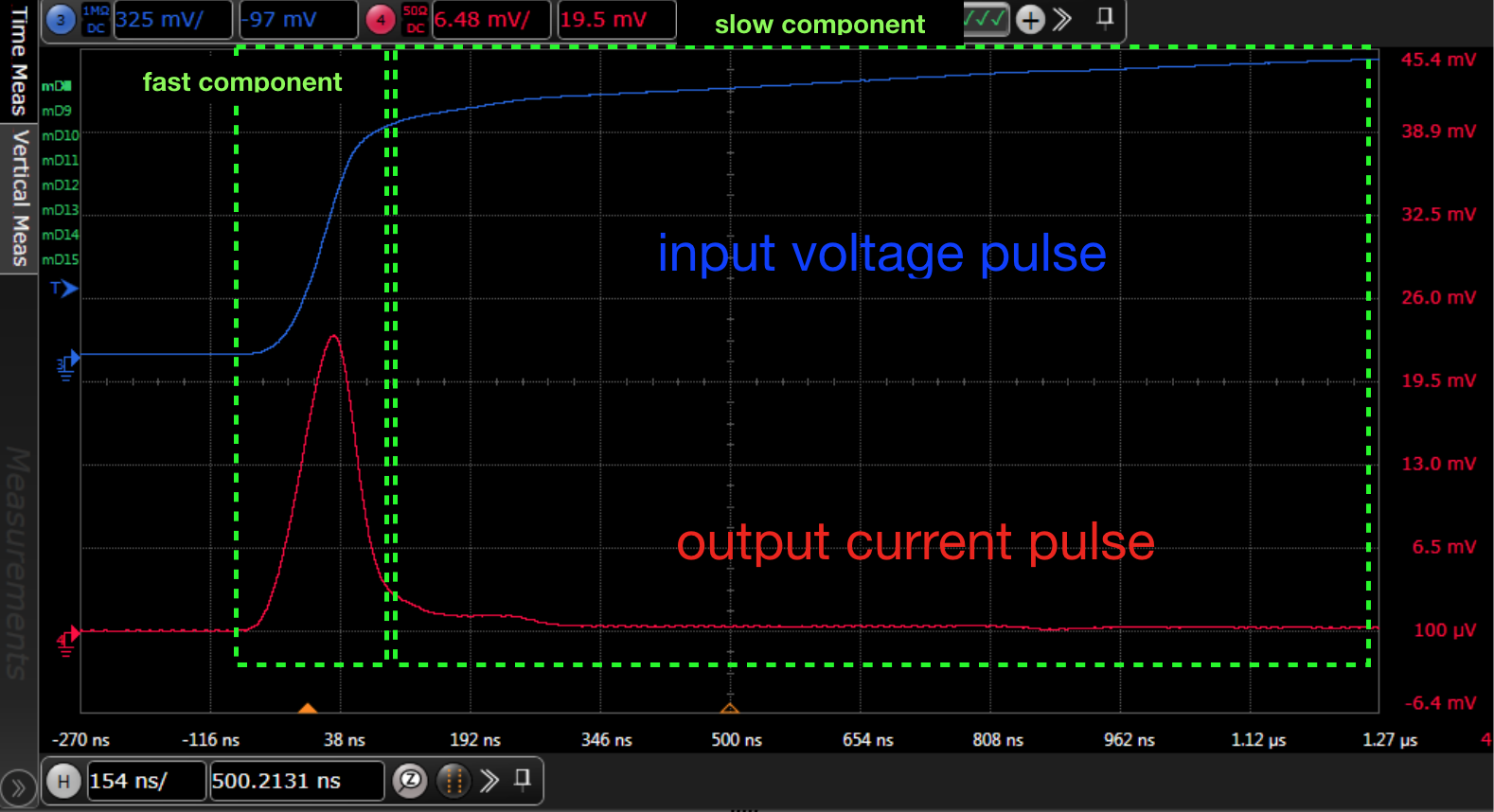}
\label{subfig: positive current}
}
\caption{(in blue) The AWG output V(t). (in red) The injector output i(t) corresponding to the sTGC emulated current signal. (\ref{subfig: complete signal}) The complete signal, including the positive current (rising edge of V(t)), followed by a negative current (falling edge of V(t)) and an artificial baseline restoration. (\ref{subfig: positive current}) Zoom-in of the positive current part of the signal. Notice the fast and slow components.
}
\label{fig: emulator signal}
\end{figure}

As described in section~\ref{sec: detector emulation}, for the detector emulation, the input signal to the current injector has to be the time integral of the detector induced current. We measured this kind of waveform from a pad of a QL1 sTGC module, having a graphite layer resistivity of $\sim$200~k$\Upomega$/$\Box$.
We used a charge sensitive preamplifier with large decay time (>ms), to ensure that any voltage change in the typical signal development timescale (up to few 100~$\upmu$s) is due to detector-induced charge movements, and it is not an artifact due to the preamplifier baseline restoration. We operated the detector in the standard CO$_2$/n-pentane (45\%/55\%) mixture at 2900~V and we recorded cosmic muon signals. The measured signals population is shown in figure~\ref{subfig: all signals}. As it can be seen, there appear to be two populations of signals, one of which has a slower rise and is due to a known cross-talk effect caused by charge propagation in the graphite layer above the pads. After implementing a more sophisticated trigger, the clean population of direct signals looks like the one in figure~\ref{subfig: fast signals}. We took the average of 350 signals. 
This signal was modified by introducing an artificial baseline restoration, and used as an arbitrary waveform input for the CAEN module described in section~\ref{sec: detector emulation}. The AWG output (V(t)) and the injector output (i(t)), are showed in figure~\ref{fig: emulator signal}. The detector signal can be divided into two regions. The fast voltage rise, corresponding to the positive current peak (figure~\ref{subfig: positive current}) is caused by the electrons/ions (mostly the ions, since the electrons are promptly collected by the wire) moving into the intense electric field region towards the wire/graphite. The different intensity of the field in space determines the change in voltage slope (the current getting smaller). The positive induced current ends when the ions land on the graphite layer. A small negative current is due to the electrons moving on the graphite surface towards the ions to neutralize them (figure~\ref{subfig: complete signal}). The intensity of the negative current and its duration depend on the resistivity of the graphite: the higher the resistivity, the smaller the current and the longer the time it persists. In our case the estimated value is $\sim$~1$\upmu$A lasting for $\sim$~130$\upmu$s. Notice that since there is no net charge collection from the anode pad, the induced current signal is purely bipolar (its integral is nil).

\newpage


\section{Efficiency formula}
\label{app: efficiency formula}

Assuming that the time between hits is random, it follows an exponential distribution $f(x)= \lambda\cdot e^{-\lambda x}$, where $x$ is the time between two hits, and $\lambda$ is the rate in Hz. We make the following approximation: in case two hits are more separated than the average intrinsic dead time D, the incurred dead time will be D. If the two hits are separated by a time $x<D$, then the dead time will be $x$. It follows that the dead time per hit is $\int_0^D x\lambda e^{-\lambda x}dx + D\int_D^\infty \lambda e^{-\lambda x}dx = \frac{1}{\lambda}(1-e^{-D\lambda}$). Multiplying by the rate we get the fractional dead time $(1-e^{-D\lambda}$). The efficiency, defined by 1-($fractional~dead~time$) is then $e^{-D\lambda}$.

\newpage    
\acknowledgments

Thanks to Siyuan Sun (University of Michigan) for the derivation of the efficiency formula, and for the collaboration during the sTGC test-beam in the CERN-GIF++ together with Petr Teterin (NRNU MEPhI), Liang Guan (University of Michigan) and Gerardo Vasquez (University of Vicrtoria). 
\\Thanks to the ATLAS-NSW project for allowing the use of test-beam data.
\\Thanks to Shay Hacohen-Gourgy (Technion) and his group for the technical support, and to Interlligent RF \& Microwave Solutions for providing the Tektronix AWG and technical support.
This research was supported by Israel Science Foundation (ISF) grant numbers (2323/18 and 1638/18) and the US-Israel Binational Science Foundation (BSF) grant number (2018360). 

\bibliographystyle{elsarticle-num}                 
\bibliography{bibliography.bib}

\end{document}